# Velocity Gauge Potentials in Electrodynamics

D. V. Giri [1], Frederick M. Tesche [2], and Michael A. Morgan [3]

*Abstract*—**Vector and scalar potentials are used for convenience in solving boundary value problems involving electromagnetic (EM) fields. The potentials are made unique by choosing a non-unique gauge relationship. The most commonly used gauges are those named for Lorenz and Coulomb, both of which may be defined as special cases of what is termed the velocity gauge, or *v*-gauge. This generalized gauge is not usually taught to students of electrodynamics. In this paper, we review properties of the velocity gauge, including EM field invariance, and demonstrate its application via an example.**

## I. INTRODUCTION

VECTOR and scalar potentials are commonly introduced to students as a mechanism to compute electromagnetic fields from specified current and charge densities, $\vec{J}(\vec{r},t)$ and $\rho(\vec{r},t)$. As part of the derivation, there comes a point where a defining relationship, termed a gauge, needs to be established between the vector and scalar potentials. Jackson [1] states "It seems necessary from time to time to show that the electric and magnetic fields are independent of the choice of gauge for the potentials."

Two commonly employed gauge relationships are named for Ludvig Valentin Lorenz and Charles-Augustin de Coulomb. Less well known, even to most antenna engineers, is that these two gauge selections are special cases of a generalized velocity gauge (*v*-gauge) [1-5], characterized by a velocity parameter *v*. The Lorenz and Coulomb cases respectively result by selecting $v = c$ and $v = \infty$ in the *v*-gauge relationship between scalar and vector potentials.

The use of gauges prompts several truly relevant questions. If one uses two different values of *v* in the *v*-gauge are the resultant electromagnetic fields the same? More generally, is gauge invariance a sacred principle in classical EM? There has been some controversy regarding gauge invariance and uniqueness. For example, Engelhardt [6] has claimed that Maxwell's equations have non-unique solutions.

In this paper, we will show that gauge invariance is <u>not</u> violated in classical electrodynamics for the continuous range of potentials encompassed in the *v*-gauge. Furthermore, we will see that the parameter *v* need not be limited to any physical range, such as *c* to infinity, but can take on any value from $-\infty < v < \infty$ except $v = 0$. If we set physical considerations aside the mathematical range of *v* can even be extended to all numbers in the complex plane, excluding the origin. Thus, the *v*-gauge provides a doubly-infinite set of gauge functions for determining the EM fields.

As an aside, it should be noted that under the conditions of classical electrodynamics, where currents and charges are assumed to be well defined functions of space and time, there is no absolute need to employ potentials. As shown in [7] and many other references it is possible to solve Maxwell's equations directly to yield integral formulas for fields in terms of sources. It is also possible to directly solve Maxwell's equations, at least approximately, via finite-element and finite-difference methods, without recourse to potentials. In fact most commercial-off-the-shelf (COTS) EM software directly computes electromagnetic fields without using potentials. At least one exception to this is the COMSOL Magnetic and Electric Fields interface in the quasi-static AC/DC Module, which solves for both scalar and vector potentials.

The first author of this paper was a student of the iconic Prof. R. W. P. King [8], who taught him, "Who needs potentials?" Although they may not be strictly needed, potentials do form a powerful and widely used addition to the set of mathematical tools employed in electrodynamics. With this in mind we will proceed to derive and apply field computation formulas using velocity gauge potentials.

This paper is organized as follows. In section II, we describe the velocity gauge using SI units. Much of the mathematical approach in Section II is based on the CGS unit formulation in Jackson's technical report [1]. In section III, we present an example application of *v*-gauge potentials to calculate the EM fields radiated by a time-harmonic Hertzian dipole. This appears to be an original effort, not found in the literature. As will be demonstrated, the scalar potential will have an apparent velocity determined by the selection of *v* in the velocity gauge. The vector potential will have a contribution which propagates with velocity *c* and another portion with apparent velocity *v*. Although the potentials will appear to be individually non-causal (unless the Lorenz gauge is selected), the electric and magnetic fields computed using these potentials will remain causal and will be invariant to the gauge's apparent velocity *v*. Summarizing comments are offered in section IV, followed by appendices showing detailed derivations, acknowledgments and listed references.

## II. THE VELOCITY GAUGE

Consider a volumetric distribution of time-varying current and charge as shown in Fig.1, with source point given by $\vec{r}\,'$. We are interested in determining the electromagnetic fields at locations outside of the source region denoted by $\vec{r}$ or in spherical coordinates $(r, \theta, \phi)$. Fields will satisfy Maxwell's equations,

$$\nabla \times \vec{E}(r,t) = -\frac{\partial \vec{B}(r,t)}{\partial t} \tag{1a}$$

$$\nabla \times \vec{H}(\vec{r},t) = \vec{J}(\vec{r},t) + \frac{\partial \vec{D}(\vec{r},t)}{\partial t} \tag{1b}$$

$$\nabla \cdot \vec{B}(\vec{r},t) = 0 \tag{1c}$$

$$\nabla \cdot \vec{D}(\vec{r},t) = \rho(\vec{r},t) \tag{1d}$$

with $\vec{B} = \mu_o \vec{H}$ and $\vec{D} = \varepsilon_o \vec{E}$ in free space, with respective permeability and permittivity denoted by $\mu_o$ and $\varepsilon_o$.

[1] D. V. Giri is with Pro-Tech, 410 Washington Street, # 1 Wellesley Hills, MA 02481-6209 and with the Dept. of Electrical and Computer Engineering, University of New Mexico, Albuquerque, NM.
[2] F. M. Tesche is an Electromagnetics Consultant living in Lakeville, CT.
[3] M. A. Morgan is an Emeritus Distinguished Professor of Electrical & Computer Engineering at the Naval Postgraduate School, Monterey, CA.

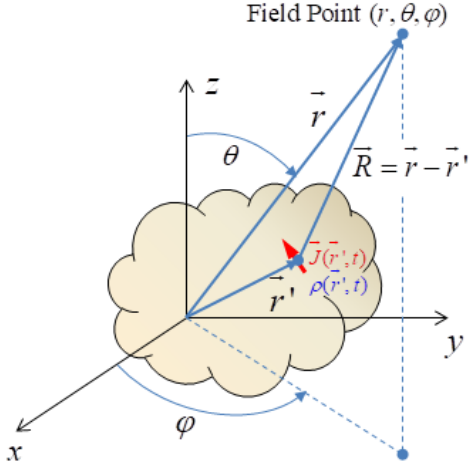

Fig. 1. A current and charge distribution producing EM fields at $\vec{r}$.

When the right side of (1a) is zero, or negligible for quasi-static and electrostatic problems, the E-field may be represented using the gradient of a scalar potential, the same function used to describe voltage in circuits. Having a single scalar potential function to represent three components of the vector E-field simplifies the solution.

Extending the use of potentials to solve the time dependent Maxwell's equations in (1a) to (1d), or to their frequency-domain versions, originates with (1c): $\nabla \cdot \vec{B}(\vec{r},t) = 0$. This infers the absence of magnetic monopoles, which means that magnetic flux lines close on themselves, as opposed to terminating on magnetic charges. Since this divergence of $\vec{B}(\vec{r},t)$ vanishes, one can write the field as the curl of a vector potential,

$$\vec{B}(\vec{r},t) = \nabla \times \vec{A}(\vec{r},t) \quad . \tag{2}$$

The consequence of (2) is that one is free to choose the divergence of the vector potential $\nabla \cdot \vec{A}(\vec{r},t)$ to be any convenient scalar function we wish and there will be no effect on the B-field.

Following a somewhat standard approach, we substitute (2) into (1a) and we note that the curl of the gradient of a scalar function, namely the scalar potential, is zero. This allows us to represent the electric field using both the scalar and vector potentials $\phi(\vec{r},t)$ and $\vec{A}(\vec{r},t)$ as

$$\vec{E}(\vec{r},t) = -\nabla \phi(\vec{r},t) - \frac{\partial \vec{A}(\vec{r},t)}{\partial t} \quad . \tag{3}$$

It is clear that if one performs the curl operation on both sides of (3) we will recover (1a). The derivation of (3) also indicates that there are an infinite number of pairs of scalar and vector potentials that generate the electric and magnetic fields using (3) and (2). To show this, we recall that the curl of the gradient of a scalar function is zero: $\nabla \times \left( \nabla \chi(\vec{r},t) \right) = 0$. Using this result we can transition from one pair of vector and scalar potentials $\left\{ \vec{A}(\vec{r},t), \phi(\vec{r},t) \right\} \rightarrow \left\{ \vec{A}'(\vec{r},t), \phi'(\vec{r},t) \right\}$ to another via

$$\vec{A}'(\vec{r},t) = \vec{A}(\vec{r},t) + \nabla \chi(\vec{r},t) \tag{4a}$$

$$\phi'(\vec{r},t) = \phi(\vec{r},t) - \frac{\partial \chi(\vec{r},t)}{\partial t} \quad . \tag{4b}$$

EM field invariance to this transition of potentials can be shown directly using

$$\begin{aligned} \vec{B}'(\vec{r},t) &= \nabla \times \vec{A}'(\vec{r},t) \\ &= \nabla \times \vec{A}(\vec{r},t) + \nabla \times \left( \nabla \chi(\vec{r},t) \right) \\ &= \nabla \times \vec{A}(\vec{r},t) = \vec{B}(\vec{r},t) \end{aligned} \tag{5a}$$

and

$$\begin{aligned} \vec{E}'(\vec{r},t) &= -\nabla \phi'(\vec{r},t) - \frac{\partial \vec{A}'(\vec{r},t)}{\partial t} \\ &= -\nabla \left( \phi(\vec{r},t) - \frac{\partial \chi(\vec{r},t)}{\partial t} \right) - \frac{\partial}{\partial t} \left( \vec{A}(\vec{r},t) + \nabla \chi(\vec{r},t) \right) \\ &= -\nabla \phi(\vec{r},t) - \frac{\partial \nabla \chi(\vec{r},t)}{\partial t} - \frac{\partial \vec{A}(\vec{r},t)}{\partial t} + \frac{\partial \nabla \chi(\vec{r},t)}{\partial t} \\ &= -\nabla \phi(\vec{r},t) - \frac{\partial \vec{A}(\vec{r},t)}{\partial t} = \vec{E}(\vec{r},t) \quad . \end{aligned} \tag{5b}$$

The scalar function $\chi(\vec{r},t)$ is termed the "gauge function" for the potentials and has units of Tesla-m². This function is used to convert pairs of vector and scalar potentials from one gauge to another via (4a) and (4b), while keeping the same electric and magnetic fields [1].

The coupled partial differential equations for the scalar and vector potentials are found as follows:

$$\nabla^2 \phi(\vec{r},t) + \frac{\partial}{\partial t} (\nabla \cdot \vec{A}(\vec{r},t)) = -\frac{\rho(\vec{r},t)}{\varepsilon_o} \tag{6a}$$

$$\begin{aligned} \nabla^2 \vec{A}(\vec{r},t) - \frac{1}{c^2} \left( \frac{\partial^2 \vec{A}(\vec{r},t)}{\partial t^2} \right) - \nabla \left( \nabla \cdot \vec{A}(\vec{r},t) + \frac{1}{c^2} \frac{\partial \phi(\vec{r},t)}{\partial t} \right) \\ = -\mu_o \vec{J}(\vec{r},t) \quad . \end{aligned} \tag{6b}$$

Let us examine the third term on the left side of (6b), recalling that we have yet to set the value of $\nabla \cdot \vec{A}(\vec{r},t)$. The following choice forms the essence of the velocity gauge,

$$\nabla \cdot \vec{A}_v(\vec{r},t) = -\frac{1}{v^2} \frac{\partial \phi_v(\vec{r},t)}{\partial t} \tag{7}$$

with *generalized* speed $v$ defined as a variable parameter. Two specific choices for $v$, namely $v = c$ and $v \to \infty$ lead to the respective definitions of the commonly employed Lorenz and Coulomb gauges, denoted by subscripts $L$ and $C$,

$$\nabla \cdot \vec{A}_L(\vec{r}, t) = -\frac{1}{c^2} \frac{\partial \phi_L(\vec{r}, t)}{\partial t} \quad \text{(Lorenz gauge)} \quad (8a)$$

$$\nabla \cdot \vec{A}_C(\vec{r}, t) = 0 \qquad \text{(Coulomb gauge). (8b)}$$

Continuing with the *v*-gauge formulation, we substitute (7) into (6a), resulting in the wave equation for the scalar potential

$$\nabla^2 \phi_v(\vec{r}, t) - \frac{1}{v^2} \frac{\partial^2 \phi_v(\vec{r}, t)}{\partial t^2} = -\frac{\rho(\vec{r}, t)}{\varepsilon_o} \quad (9)$$

Using the standard Green's function solution for the inhomogeneous wave equation [1] we find that

$$\phi_v(\vec{r}, t) = \frac{1}{4\pi\varepsilon_o} \iiint_V \frac{1}{R} \rho(\vec{r}', t - R/v) \, dV'. \quad (10)$$

By substituting $v = c$ or $v \to \infty$ into equations (9) and (10) we get the well-known differential equations and solutions for the scalar potentials of the Lorenz or Coulomb gauges.

To determine the vector potential for the *v*-gauge we follow the formulation in [1]. This begins by using the scalar gauge function $\chi(\vec{r}, t)$ appearing in (4b) to define differences of scalar potentials between the Lorenz and *v*-gauges,

$$\frac{\partial \chi_v}{\partial t} = \phi_L(\vec{r}, t) - \phi_v(\vec{r}, t)$$

$$= \frac{1}{4\pi\varepsilon_o} \iiint_V \frac{1}{R} \left[ \rho(\vec{r}', t - R/c) - \rho(\vec{r}', t - R/v) \right] dV'. \quad (11)$$

The remaining details of the vector potential derivation are given in Appendix A, following that given in [1], with result being

$$\vec{A}_v(\vec{r}, t) = \frac{\mu_o}{4\pi} \iiint_V \frac{1}{R} \begin{bmatrix} \vec{J}(\vec{r}', t - R/c) - \hat{R}\left(\hat{R} \cdot \vec{J}(\vec{r}', t - R/c)\right) \\ + \frac{c^2}{v^2} \hat{R}\left(\hat{R} \cdot \vec{J}(\vec{r}', t - R/c)\right) \end{bmatrix} dV'$$

$$+ \frac{1}{4\pi\varepsilon_o} \iiint_V \frac{1}{R^3} \int_{R/v}^{R/c} \tau \, d\tau \left[ 3\hat{R}\left(\hat{R} \cdot \vec{J}(\vec{r}', t - \tau)\right) - \vec{J}(\vec{r}', t - \tau) \right] dV'. \quad (12a)$$

By using the continuity equation in developing the equation above we are able to express the vector potential entirely in terms of the current density. An alternate version of (12a), also given in [1], results by retaining the charge density while computing the vector potential using integrations over both current and charge sources,

$$\vec{A}_v(\vec{r}, t) = \frac{\mu_o}{4\pi} \iiint_V \frac{1}{R} \left\{ \vec{J}(\vec{r}', t - R/c) - \hat{R} c \rho(\vec{r}', t - R/c) \right.$$

$$\left. + \hat{R} \frac{c^2}{v} \rho(\vec{r}', t - R/v) + \frac{\hat{R}}{R} c^2 \int_{R/v}^{R/c} d\tau \, \rho(\vec{r}', t - \tau) \right\} dV'. \quad (12b)$$

As a check of these results, if we set $v = c$ in either (12a) or (12b) there is cancellation of all integration terms but one, yielding the well-known Lorenz gauge expression

$$\vec{A}_L(\vec{r}, t) = \frac{\mu_o}{4\pi} \iiint_V \frac{1}{R} \vec{J}(\vec{r}', t - R/c) \, dV', \quad (13)$$

Pairing equation (10) with either (12a) or (12b) completes the formulation of the velocity gauge potentials.

The final step in our derivation is to employ the velocity gauge potentials to express the electromagnetic fields, using (2) for $\vec{B}$ and (3) for $\vec{E}$. Starting with the electric field, we calculate the two terms in (3) separately, using (10) for the first term

$$-\nabla \phi_v(\vec{r}, t)$$

$$= \frac{1}{4\pi\varepsilon_o} \iiint_V \frac{1}{R} \left[ \frac{\hat{R}}{R} \rho(\vec{r}', t - R/v) + \hat{R} \frac{1}{v} \frac{\partial}{\partial t} \rho(\vec{r}', t - R/v) \right] dV' \quad (14)$$

The second term of (3) is found using (12b)

$$-\frac{\partial \vec{A}_v(\vec{r}, t)}{\partial t} =$$

$$\frac{\mu_o}{4\pi} \iiint_V \frac{1}{R} \begin{bmatrix} -\frac{\partial \vec{J}(\vec{r}', t - R/c)}{\partial t} + \hat{R} c \frac{\partial \rho(\vec{r}', t - R/c)}{\partial t} \\ -\hat{R} \frac{c^2}{v} \frac{\partial \rho(\vec{r}', t - R/v)}{\partial t} \\ -\frac{\hat{R}}{R} c^2 \int_{R/v}^{R/c} d\tau \frac{\partial \rho(\vec{r}', t - \tau)}{\partial t} \end{bmatrix} dV' \quad (15a)$$

The $\partial \rho / \partial t$ term in the integral of (15a) is equal to $-\partial \rho / \partial \tau$, which allows evaluation of the $\tau$ - integral in the brackets,

$$-\frac{\partial \vec{A}_v(\vec{r}, t)}{\partial t} =$$

$$\frac{\mu_o}{4\pi} \iiint_V \frac{1}{R} \begin{bmatrix} -\frac{\partial \vec{J}(\vec{r}', t - R/c)}{\partial t} + c \hat{R} \frac{\partial \rho(\vec{r}', t - R/c)}{\partial t} \\ +\frac{\hat{R}}{R} c^2 \rho(\vec{r}', t - R/c) \\ -\frac{\hat{R}}{R} c^2 \rho(\vec{r}', t - R/v) - \hat{R} \frac{c^2}{v} \frac{\partial \rho(\vec{r}', t - R/v)}{\partial t} \end{bmatrix} dV'. \quad (15b)$$

By adding (14) to (15b), and noting that $c^2 = 1/(\mu_o \varepsilon_o)$, we end up canceling all integration terms which propagate at the non-causal velocity $v$. The result is

$$\vec{E}_v(\vec{r},t) = \vec{E}_L(\vec{r},t) =$$

$$\frac{\mu_o}{4\pi} \iiint_V \frac{1}{R} \begin{bmatrix} -\dfrac{\partial \vec{J}(\vec{r}\,',t-R/c)}{\partial t} + c\hat{R}\,\dfrac{\partial \rho(\vec{r}\,',t-R/c)}{\partial t} \\ + c^2\,\dfrac{\hat{R}}{R}\,\rho(\vec{r}\,',t-R/c) \end{bmatrix} dV' \quad (16)$$

This expression is the same as found by substituting the Lorenz gauge potentials of (13) and (10), using $v = c$, into (5b) which shows that the electric field found using the velocity gauge potentials is independent of the selected $v$ and is thus gauge invariant. We already knew this using (5b) but have verified the result using the $v$-gauge potentials.

We can derive the corresponding magnetic flux density by taking the curl of the $v$-gauge vector potential from (12b) as follows:

$$\vec{B}_v(\vec{r},t) = \nabla \times \vec{A}_v(\vec{r},t) =$$

$$\frac{\mu_o}{4\pi} \nabla \times \iiint_V \frac{1}{R} \begin{bmatrix} \vec{J}(\vec{r}\,',t-R/c) - \hat{R}c\,\rho(\vec{r}\,',t-R/c) \\ + \hat{R}\dfrac{c^2}{v}\,\rho(\vec{r}\,',t-R/v) \\ + \dfrac{\hat{R}}{R}c^2 \displaystyle\int_{R/v}^{R/c} d\tau\,\rho(\vec{r}\,',t-\tau) \end{bmatrix} dV'. \quad (17)$$

The curl, operating on the unprimed coordinates, may be taken inside the primed coordinate volume integral. Note that each of the terms containing the charge density has the vector form $\hat{R}\,f(\vec{r},\vec{r}\,',t)$, where $\hat{R} = (\vec{r}-\vec{r}\,')/R$. It can be shown that when the curl operates on the $\vec{r}$ field point of such a form the result is zero: $\nabla \times \left[\hat{R}\,f(\vec{r},\vec{r}\,',t)\right] = \vec{0}$. Thus, none of the $q(t) = (I_0/\omega)\sin\omega t$ terms containing the charge density produce contributions. The result is that the $B$ field from the $v$-gauge depends only on the current density and is given by

$$\vec{B}_v(\vec{r},t) = \frac{\mu_o}{4\pi}\,\nabla \times \left[\iiint_v \frac{1}{R}\vec{J}(\vec{r}\,',t-R/c)\,dV'\right] = \vec{B}_L(\vec{r},t). \quad (18)$$

This is the same result as obtained by substituting the Lorenz gauge $\vec{A}_L(\vec{r},t)$ in (13) into (4b). As expected from (5a), this confirms the invariance of the B-field using the $v$-gauge.

We take the field point curl inside of the integral of (18) using the product rule for curls,

$$\nabla \times \left[\vec{J}(\vec{r}\,',\tau)/R\right] = (1/R)\,\nabla \times \vec{J}(\vec{r}\,',\tau) + \nabla(1/R) \times \vec{J}(\vec{r}\,',\tau)$$

resulting in

$$\vec{B}_v(\vec{r}\,',t) =$$

$$\frac{\mu_o}{4\pi} \iiint_v \frac{1}{R}\,\nabla \times \vec{J}(\vec{r}\,',t-R/c) - \frac{\hat{R}}{R^2} \times \vec{J}(\vec{r}\,',t-R/c)\,dV'. \quad (19)$$

where it is important to recognize that

$$\nabla \times \vec{J}(\vec{r}\,',t-R/c) =$$

$$\frac{\partial \vec{J}(\vec{r}\,',t-R/c)}{\partial t'} \times \nabla\left(t-\frac{R}{c}\right) = \frac{1}{c}\frac{\partial \vec{J}(\vec{r}\,',t-R/c)}{\partial t'} \times \hat{R}. \quad (20)$$

Substituting (20) into (19), yields

$$\vec{B}_v(\vec{r},t') = \vec{B}_L(\vec{r},t')$$

$$= \frac{\mu_o}{4\pi} \iiint_V \left[\frac{1}{R^2}\vec{J}(\vec{r}\,',t-R/c) + \frac{1}{cR}\frac{\partial}{\partial t'}\vec{J}(\vec{r}\,',t-R/c)\right] \times \hat{R}\,dV'.$$

Equations (16) and (21) are termed the "Jefimenko" expressions, [9]. These appear in [1] and are described online at Wikipedia, which gives additional citations.

In concluding this section, we emphasize that while the potentials depend on the gauge used the electromagnetic fields must be gauge invariant for any gauge used, as was shown by substituting (4a) and (4b) into (5a) and (5b).

### III. VELOCITY GAUGE EXAMPLE

We will consider the Hertz dipole shown in Fig. 2. A time harmonic variation of a current element is assumed with $I(t) = I_o \cos\omega t$ flowing in an infinitesimally thin conductor along the $z$-axis over a differential length $-d\ell/2 \le z' \le d\ell/2$.

Due to charge continuity, $I(t) = dq/dt$, equal and opposite time-varying charges $\pm q(t)$ exist at the ends of the differential filament, as shown in the figure. Simple integration yields $q(t) = (I_0/\omega)\sin\omega t$.

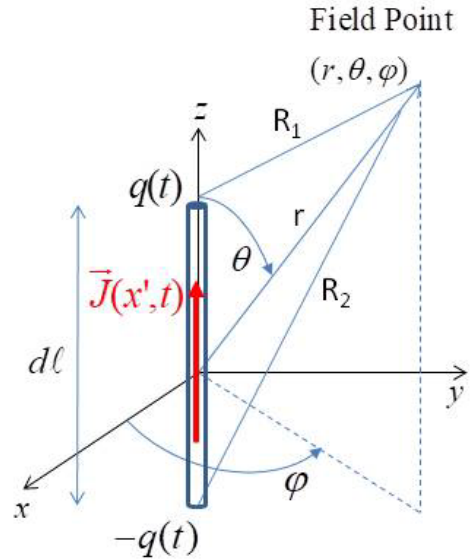

Fig. 2. Hertz dipole in a spherical coordinate system.

The Hertz dipole appears in many antenna texts, usually serving as an example for field calculation using the Lorenz gauge potentials [10–11]. Employment of the velocity gauge potentials to compute the Hertz dipole fields could not be found in the literature so is performed here in detail, perhaps for the first time.

The Hertz dipole current density may be expressed using Dirac delta functions and Heaviside unit step functions as

$$\vec{J}\left(\vec{r}\,',t\right)=\hat{z}\,I(t)\,\delta(x')\,\delta(y')\left[u\left(z'+d\ell/2\right)-u\left(z'-d\ell/2\right)\right]. \quad (22)$$

The resultant charge density is given by

$$\rho(\vec{r}\,',t)=q(t)\,\delta(x')\,\delta(y')\left[\delta\left(z'-d\ell/2\right)-\delta\left(z'+d\ell/2\right)\right]. \quad (23)$$

The velocity gauge scalar potential as computed using (10) with the charge density in (23) where $q(t)=\left(I_o/\omega\right)\sin\omega t$. The result is carefully derived in detail in Appendix B,

$$\phi_v(r,\theta,t)=\frac{1}{4\pi\varepsilon_o}\iiint\limits_V\frac{1}{R}\rho(\vec{r}\,',\,t-R/v)\,dV'$$

$$=\frac{I_o d\ell \cos\theta}{4\pi\varepsilon_o}\left[\frac{1}{v\,r}\cos\omega\left(t-r/v\right)+\frac{1}{\omega\,r^2}\sin\omega\left(t-r/v\right)\right] \quad (24)$$

Notice that this scalar potential propagates away from the dipole with velocity $v$ and has portions that separately track the time variation of current and charges on the dipole.

The Lorenz case is found simply by setting $v=c$ in (24) while the Coulomb gauge case uses $v\to\infty$, giving

$$\phi_c(r,\theta,t)\doteq\frac{I_o d\ell \cos\theta}{4\pi\varepsilon_o\,\omega\,r^2}\sin\left(\omega t\right) \quad (25)$$

where the contribution involving the current vanishes, leaving a result that is identical to the scalar potential for a static dipole. The Coulomb gauge scalar potential tracks the charge variation in real-time at any distance without *any* time delay.

To compute the $v$-gauge vector potential we will apply (4a)

$$\vec{A}_v(r,\theta,t)=\vec{A}_L(r,\theta,t)+\nabla\chi_v(r,\theta,t) \quad (26)$$

to transition from the Lorenz gauge vector potential. The Lorenz potential is evaluated in Appendix B using (13) with current density in (22), where $I(t)=I_o\sin\omega t$. The result is

$$\vec{A}_L(\vec{r},t)=\frac{\mu_o}{4\pi}\iiint\limits_V\frac{1}{R}\vec{J}\left(\vec{r}\,',\,t-R/c\right)dV'$$

$$=\hat{z}\frac{\mu_o I_0 d\ell}{4\pi r}\cos\omega\left(t-r/c\right). \quad (27)$$

As expected, the Lorenz gauge vector potential follows only the time-variation of the current while propagating away from the dipole at velocity $c$ with inverse distance decrease.

Continuing with use of (26), we need to evaluate $\nabla\chi_v(r,\theta,t)$. This is done, first by applying (4b)

$$\frac{\partial\chi_v(r,\theta,t)}{\partial t}=\phi_L(r,\theta,t)-\phi_v(r,\theta,t). \quad (28)$$

where both $\phi_v(r,\theta,t)$ and $\phi_L(r,\theta,t)$ are obtained from (24), the latter using $v=c$. Details of the calculation are given in Appendix B, with answer

$$\vec{A}_v(r,\theta,t)=\frac{I_o d\ell}{4\pi\varepsilon_o\,\omega}\left[\hat{r}\,R_v(r,t)\cos\theta+\hat{\theta}\,T_v(r,t)\sin\theta\right] \quad (29a)$$

where

$$R_v(r,t)=-\frac{2}{c\,r^2}\sin\omega(t-r/c)+\frac{2}{\omega r^3}\cos\omega(t-r/c)$$

$$+\frac{2}{v r^2}\sin\omega(t-r/v)+\frac{\omega}{v^2 r}\cos\omega(t-r/v)$$

$$-\frac{2}{\omega r^3}\cos\omega(t-r/v) \quad (29b)$$

$$T_v(r,t)=-\frac{\omega}{c^2 r}\cos\omega(t-r/c)-\frac{1}{c\,r^2}\sin\omega(t-r/c)$$

$$+\frac{1}{\omega r^3}\cos\omega(t-r/c)+\frac{1}{v r^2}\sin\omega(t-r/v)$$

$$-\frac{1}{\omega r^3}\cos\omega(t-r/v). \quad (29c)$$

This result can also be obtained by substituting (22) and (23) into (12b) followed by careful integration. However, the procedure shown in Appendix B, starting from (26), illustrates use of the gauge function to transition the vector potential from one gauge to another.

It is interesting to note that $\vec{A}_v(r,\theta,t)$ has parts which propagate at $c$, like the Lorenz gauge, while other portions propagate with arbitrary velocity $v$. If $v$ is set to $c$ in (29b) and (29c) all terms cancel except those which vary as $r^{-1}$. By using $\mu_0\,\varepsilon_0=c^{-2}$ and $\hat{z}=\hat{r}\cos\theta-\hat{\theta}\sin\theta$ the result for $v=c$ in (29) reduces to that of the Lorenz gauge in (27).

Before proceeding with field calculations for the Hertz dipole we will further investigate the behavior of the velocity gauge potentials in this example. To do this we will consider the phasor forms of the potentials, where the conversions from complex phasor to time-harmonic function are given by

$$f(\vec{r},t)=\operatorname{Re}\left[F(\vec{r},\omega)\,e^{j\omega t}\right] \quad (30)$$

Using (30) the phasor representation of $\phi_v(r,\theta,t)$ in (24) is found to be

$$\phi_v(r,\theta,t)=\operatorname{Re}\left[\frac{I_o Z_0\,kd\ell\cos\theta}{4\pi}\Phi(kr,v)\,e^{j\omega t}\right] \quad (31a)$$

where the dimensionless distance dependent portion of the phasor function is

$$\Phi(kr, v) = \frac{1}{kr}\left[\frac{c}{v} + \frac{1}{jk\,r}\right]e^{-jk_v r} \qquad (31b)$$

with free-space wave number $k = \omega / c$ and velocity gauge wave number $k_v = \omega / v$.

The magnitude of $\left|\Phi(kr, v)\right|$ for seven values of $v$ are plotted in Fig. 3 versus dimensionless $kr$ to demonstrate the dependence on assumed velocity $v$. As seen in (24) or (31b), all portions of the scalar potential propagate at $v$. For the Lorenz gauge case, with $v = c$ (red line in Fig. 3), the near-field to far-field transition for the Hertz dipole occurs in the vicinity of $kr = 1$. In the near-field, with $kr \ll 1$, the scalar potential varies with the inverse square of radial distance while in the far-field, with $kr \gg 1$, the scalar potential varies with the inverse of the radial distance.

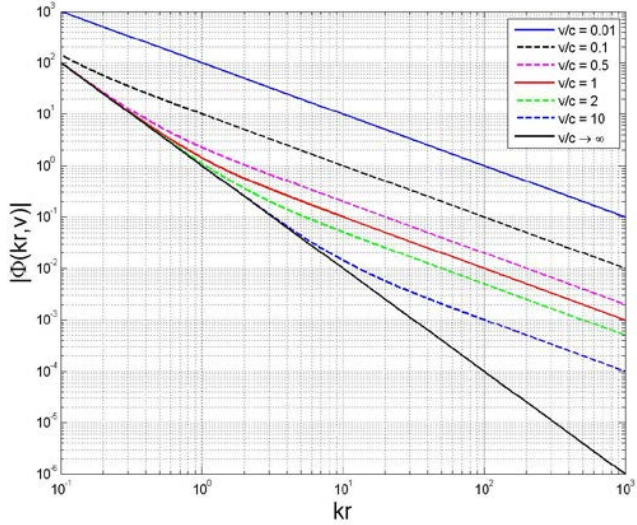

Fig. 3  Magnitude of the scalar potential phasor factor $\left|\Phi\right|$

The same form of near- to far-field transition occurs for the $v \neq c$ cases in the vicinity of $kr = v/c$. See for example the dashed blue line case in Fig. 3 where $v = 10c$.

The Coulomb gauge case, with $v / c \to \infty$ is plotted by the black line in Fig. 3, where the actual value used is $v / c = 10^9$. The resultant scalar potential has an inverse-squared $r^{-2}$ quasi-static radial variation for all log-log plotted values in the range of $kr$ shown. This results by letting $c / v \to 0$ in the phasor of (31b) or by considering the time-harmonic Coulomb gauge potential in (25).

At the other extreme with $v / c = .01$ (solid blue line), the scalar potential propagates away from the dipole at $.01c$ and exhibits an inverse-distance $r^{-1}$ far-field behavior over the plotted range of $kr$.

Assuming we are considering a non-Lorenz velocity gauge, how can the scalar potential, propagating at $v \neq c$, possibly contribute to the $v = c$ propagating electric field via (5b)? To address this question we need to consider the behavior of the vector potential components.

The phasor representation for the velocity gauge vector potential in (29) is given by

$$\vec{A}_v(r, \theta, t) = \mathrm{Re}\left\{\frac{I_o Z_o\, k d\ell}{4\pi\, c}\left[\hat{r}\,\mathrm{A}_r(kr, v)\cos\theta + \hat{\theta}\,\mathrm{A}_\theta(kr, v)\sin\theta\right]e^{j\omega t}\right\}$$

$$(32a)$$

where the dimensionless $r$-dependent factors for the phasor components are

$$\mathrm{A}_r = \left[\frac{2j}{(k\,r)^2} + \frac{2}{(kr)^3}\right]e^{-jkr} + \left[\frac{(c/v)^2}{kr} - \frac{2j(c/v)}{(k\,r)^2} - \frac{2}{(kr)^3}\right]e^{-jk_v r} \quad (32b)$$

and

$$\mathrm{A}_\theta = \left[-\frac{1}{kr} + \frac{j}{(kr)^2} + \frac{1}{(kr)^3}\right]e^{-jkr} - \left[\frac{j(c/v)}{(kr)^2} + \frac{1}{(kr)^3}\right]e^{-jk_v r} \quad (32c)$$

Magnitudes of $\left|\mathrm{A}_r(kr, v)\right|$ and $\left|\mathrm{A}_\theta(kr, v)\right|$ are respectively plotted in Figs. 4a and 4b for seven values of $v$. As seen in (32b) and (32c) we expect to have contributions to the plotted functions which vary as $(kr)^{-1}$, $(kr)^{-2}$ and $(kr)^{-3}$. Further, in the non-Lorenz case, where $v \neq c$, each vector component has two bracketed terms, one with phase factor $e^{-jkr}$ which propagates at $v = c$ and the other with factor $e^{-jk_v r}$ which propagates at $v \neq c$. The time-harmonic versions of these terms also appear in formulas of (29b) and (29c).

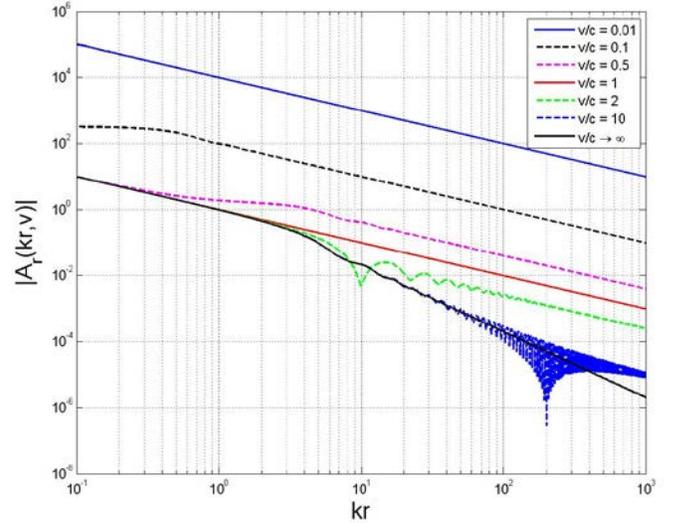

Fig. 4a.  Magnitude of the vector potential phasor factor $\left|\mathrm{A}_r\right|$.

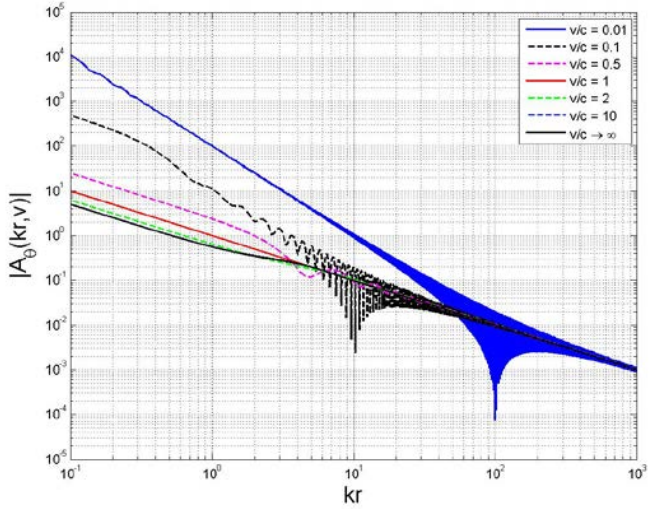

Fig. 4b. Magnitude of the vector potential phasor factor $\left|A_\theta\right|$.

The exception to this is the Lorenz gauge case, with $v = c$, where $k_v \approx k$ so the bracketed terms each propagate with velocity $c$ and the inverse-square and inverse-cubed terms will cancel, leaving just the far-field $(kr)^{-1}$ term. This result is given in the time-harmonic equation (27) and appears as the straight red line in the log-log plots of Figs. 4a and 4b.

The pairs of square-bracketed terms in each of the two equations (32b) and (32c) propagate at different velocities and thus interact as do two sinusoids having different spatial frequencies. This interaction results in a strong interference at radial distances where the magnitudes of the respective square-bracketed terms are comparable.

Inspecting (32b) and equating magnitudes of the first terms within each of the two square brackets we see that the interference between the terms in $A_r$ will be the strongest in the vicinity of $kr \approx 2\left(v/c\right)^2$. This formula predicts that in Fig 4a interference patterns will occur near $kr \approx 200$ when $v/c$ =10 and near $kr \approx 8$ when $v/c$ =2. It also predicts interference patterns for $v/c < 1$. These subluminal cases appear less rapid and smaller in amplitude due to both smaller $kr$ values and the scaling of the log-log plot. Interference for the $v/c$=.01 case is to the left of $kr$ values plotted so is not shown.

Considering $A_\theta$ in (32c), and again equating magnitudes of the first terms in the two square brackets, it is found that the interference will be strongest in the vicinity of $kr \approx c/v$. This predicts the observed oscillations in Fig. 4b with $v/c$ = 0.5, 0.1 and 0.01. As with Fig 4a, the log-log plot stretches out and reduces the apparent amplitude of displayed interference patterns the appear for smaller $kr$ values ($v/c > 1$ in this case). The $v/c = 10$ case appears to be missing but is just hidden behind the $v/c \to \infty$ plot for the Coulomb gauge.

We will not turn our attention back to computation of the electromagnetic fields using the velocity gauge. Starting with the B-field, we judiciously insert the vector potential from (26) into (2) and perform the curl operation. Because the curl operation on the gradient of any scalar function is zero, including $\nabla \times \nabla \chi_v(r, \theta, t) = 0$, we find that

$$\vec{B}_v(r, \theta, t) = \nabla \times \vec{A}_L(r, \theta, t) + \nabla \times \nabla \chi_v(r, \theta, t) = \nabla \times \vec{A}_L(r, \theta, t) = \vec{B}_L(r, \theta, t)$$

$$= \frac{\mu_o I_o \, d\ell \sin\theta}{4\pi} \hat{\varphi} \left[\frac{1}{r^2}\cos\omega\left(t - r/c\right) - \left(\frac{\omega}{c}\right)\frac{1}{r}\sin\omega\left(t - r/c\right)\right]. \tag{33}$$

This approach shows, almost trivially, that the B-field found via the velocity gauge is invariant to the chosen $v$ and equal to the result using the Lorenz gauge. Of course we had already shown this for an arbitrary gauge function (5a). Had we instead decided to evaluate the curl of $\vec{A}_v(r, \theta, t)$ given in (29) instead of that in (26) the result would still be that shown in (33). This was done as an exercise but is not shown here.

The electric field can be found by integrating (1b) outside of the source, where $\vec{J}(\vec{r}, t) = 0$, to give

$$\vec{E}_v(r, \theta, t) = c^2 \int \nabla \times \vec{B}_v(r, \theta, t) \, dt$$
$$= c^2 \int \nabla \times \vec{B}_L(r, \theta, t) \, dt = \vec{E}_L(r, \theta, t) \tag{34}$$

This clearly demonstrates gauge-invariance. Instead, we will use the $v$-gauge vector potentials from (24) and (29) to directly evaluate the electric field using (3),

$$\vec{E}_v(\vec{r}, t) = -\nabla \phi_v(\vec{r}, t) - \frac{\partial \vec{A}_v(\vec{r}, t)}{\partial t} \tag{35}$$

The gradient of $\phi_v$ in (24) using spherical coordinates is

$$\nabla \phi_v(r, \theta, t) = \hat{r}\frac{\partial \phi_v}{\partial r} + \hat{\theta}\frac{1}{r}\frac{\partial \phi_v}{\partial \theta}$$

$$= \frac{I_o \, d\ell}{4\pi \varepsilon_o}\left[\hat{r} U_v(r, t)\cos\theta + \hat{\theta} V_v(r, t)\sin\theta\right] \tag{36a}$$

where

$$U_v(r, t) = -\frac{2}{v \, r^2}\cos\omega\left(t - r/v\right) + \frac{\omega}{v^2 \, r}\sin\omega\left(t - r/v\right)$$

$$-\frac{2}{\omega \, r^3}\sin\omega\left(t - r/v\right) \tag{36b}$$

$$V_v(r, t) = -\frac{1}{v \, r^2}\cos\omega\left(t - r/v\right) - \frac{1}{\omega \, r^3}\sin\omega\left(t - r/v\right) \tag{36c}$$

Since the electric field propagates away from the dipole at velocity $c$ we know that for if $v \neq c$ the entire $\nabla \phi_v(r, \theta, t)$ appearing in (35) must be cancelled by parts of $\partial \vec{A}_v(\vec{r}, t)/\partial t$ that also propagate at $v \neq c$. Also of concern is what happens when $v = c$. To consider this in detail, we use (29) to evaluate

$$\frac{\partial \vec{A}_v(r, \theta, t)}{\partial t} = \frac{I_o \, d\ell}{4\pi \varepsilon_o}\left[\hat{r}\frac{1}{\omega}\frac{\partial R_v(r, t)}{\partial t}\cos\theta + \hat{\theta}\frac{1}{\omega}\frac{\partial T_v(r, t)}{\partial t}\sin\theta\right] \tag{37a}$$

where

$$\frac{1}{\omega}\frac{\partial R_v(r,t)}{\partial t} = -\frac{2}{c\,r^2}\cos\omega(t-r/c) - \frac{2}{\omega\,r^3}\sin\omega(t-r/c)$$

$$+ \frac{2}{v\,r^2}\cos\omega(t-r/v) - \frac{\omega}{v^2 r}\sin\omega(t-r/v) + \frac{2}{\omega\,r^3}\sin\omega(t-r/v) \tag{37b}$$

$$\frac{1}{\omega}\frac{\partial T_v(r,t)}{\partial t} = \frac{\omega}{c^2 r}\sin\omega(t-r/c) - \frac{1}{c\,r^2}\cos\omega(t-r/c)$$

$$- \frac{1}{\omega\,r^3}\sin\omega(t-r/c) + \frac{1}{v\,r^2}\cos\omega(t-r/v) + \frac{1}{\omega\,r^3}\sin\omega(t-r/v) \tag{37c}$$

Substituting the contributions from (36) and (37) into (35) we see that the last three terms in the radial component of $D_t R_v$ in (37b), which propagating at $v$ exactly cancel all three terms which form the $U_v(r,t)$ radial component in (36b) contributed by $\nabla\phi_v(r,\theta,t)$. Likewise, the last two terms of $D_t T_v$ in (37c), which propagate at $v$, exactly cancel the two terms of $V_v(r,t)$ in the latitudinal vector component of $\nabla\phi_v(r,\theta,t)$. All remaining terms are contained in the vector potential time-derivative and propagate at $c$, giving,

$$\vec{E}_v(\vec{r},t) = -\nabla\phi_v(\vec{r},t) - \frac{\partial \vec{A}_v(\vec{r},t)}{\partial t}$$

$$= -\nabla\phi_L(\vec{r},t) - \frac{\partial \vec{A}_L(\vec{r},t)}{\partial t} = \vec{E}_L(\vec{r},t) \tag{38}$$

$$= \hat{r}\,\frac{I_o\,d\ell\cos\theta}{4\pi\varepsilon_o}\left[\frac{2}{c\,r^2}\cos\omega(t-r/c) + \frac{2}{\omega\,r^3}\sin\omega(t-r/c)\right]$$

$$+ \hat{\theta}\,\frac{I_o\,d\ell\sin\theta}{4\pi\varepsilon_o}\left[\begin{array}{c}-\dfrac{\omega}{c^2 r}\sin\omega(t-r/c) + \dfrac{1}{c\,r^2}\cos\omega(t-r/c) \\[2mm] + \dfrac{1}{\omega\,r^3}\sin\omega(t-r/c)\end{array}\right]$$

The parameter $v$ can thus take on any value in the complex plane (except zero) and the electric field will remain the same. The same applies to the magnetic field.

Before concluding let us look at the frequency-domain phasor expressions for the time-harmonic electromagnetic fields. The magnetic flux density field in (33) can be written in phasor form as

$$\vec{B}_v(r,\theta,t) = \vec{B}_L(r,\theta,t) = \mathrm{Re}\left\{\hat{\varphi}\,\frac{\mu_o I_o\,d\ell\sin\theta}{4\pi}k^2 \mathrm{B}_\varphi(kr)\,e^{j\omega t}\right\} \tag{39a}$$

where the normalized radial dependence is given by

$$\mathrm{B}_\varphi(kr) = \left[\frac{1}{(kr)^2} + \frac{j}{kr}\right]e^{-jkr}. \tag{39b}$$

The corresponding electric field phasor form for (39) is

$$\vec{E}_v(r,\theta,t) = \vec{E}_L(r,\theta,t)$$

$$= \mathrm{Re}\left\{\frac{I_o\,d\ell}{4\pi}k^3 Z_o\left[\hat{r}\,\mathrm{E}_r(kr)\cos\theta + \hat{\theta}\,\mathrm{E}_\theta(kr)\sin\theta\right]e^{j\omega t}\right\} \tag{40a}$$

where $Z_0 = \sqrt{\dfrac{\mu_0}{\varepsilon_0}} = 120\pi\ \Omega$ is the wave impedance, and

$$\mathrm{E}_r(kr) = \left[\frac{2}{(kr)^2} - \frac{j2}{(kr)^3}\right]e^{-jkr} \tag{40b}$$

$$\mathrm{E}_\theta(kr) = \left[\frac{j}{kr} + \frac{1}{(kr)^2} - \frac{j}{(kr)^3}\right]e^{-jkr}. \tag{40c}$$

Figure 5 plots magnitudes of the normalized radial function components of the phasor fields shown in (39b), (40b) and (40c). These plots illustrate the differences and transitions of the radial behavior of the dipole field components in the near-zone ($kr \ll 1$) and the far-zone ($kr \gg 1$). Near to the dipole portions of both electric field components that vary as $r^{-3}$ dominate the amplitude of the vector field. This near-zone E-field has the same form as that of a $z$-directed static electric dipole, but with time-varying $\pm q(t)$ separated by $d\ell$. The corresponding near-zone B-field varies as $r^{-2}$ with behavior akin to a quasi-static current segment $I(t)$. In the far-zone, the transverse to $\hat{r}$ portions of both the electric and magnetic fields ($E_\theta$ and $H_\varphi$) vary as $r^{-1}$, with the vector fields related by $\vec{E}(r,\theta,t) \doteq c\,\vec{B}(r,\theta,t)\times\hat{r}$.

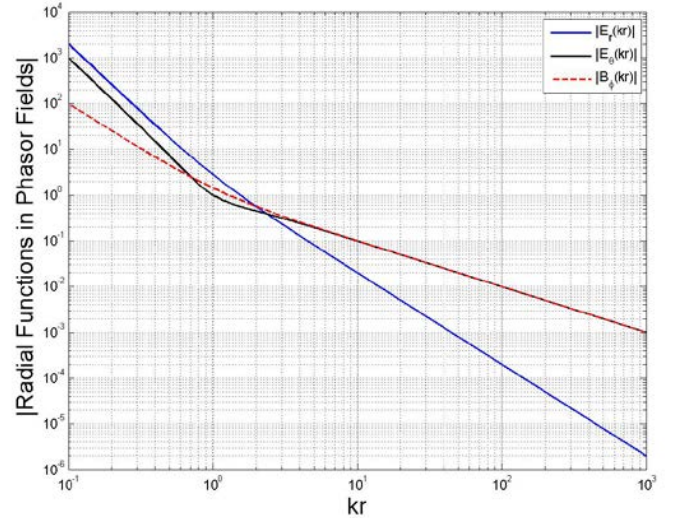

Fig. 5. Normalized EM field magnitudes (independent of $v$)

## IV. SUMMARIZING COMMENTS

In this paper, we have reviewed how the choice of gauge appears in defining the relationship between the scalar and vector potentials used to compute electromagnetic fields from sources in classical electrodynamics. Although the Lorenz and Coulomb gauge relationships are commonly taught and utilized, their generalization via the velocity gauge is not well-known. The main purpose of this paper has been to fill in this knowledge gap.

As shown here, the potentials depend on the selected gauge. In the case of the velocity gauge, with selected, $v \doteq c$ the scalar potential will propagate away from the source at the selected $v$ while a portion of the vector potential will do the same while another portion of the vector potential will remain causal, propagating at $c$. Of particular interest is the Coulomb gauge, where $v \to \infty$, producing instantaneous propagation of the entire scalar potential and a portion of the vector potential.

The velocity gauge potentials will be non-causal, and hence non-physical, unless the Lorenz gauge with $v = c$ is selected. Nonetheless, even if the potentials are non-causal the resultant electromagnetic fields will always be unique and causal. This field uniqueness is enforced explicitly in the derivation of the gauge relationships that are allowed. As shown in the example, when $v \neq c$, the non-causal contributions to the EM fields from the velocity gauge potentials are canceled.

## APPENDIX A
### Derivation of Equation (12a)

Following Jackson's derivation in [1], but using SI units, the bracketed term in (11) can be written as

$$\left[ \rho\left(\vec{r}\,',t-R/c\right) - \rho\left(\vec{r}\,',t-R/v\right) \right] = \int_{R/v}^{R/c} d\tau \frac{\partial}{\partial t} \rho\left(\vec{r}\,',t-\tau\right) \quad \text{(A1)}$$

Substituting (A1) into (11), we obtain

$$\frac{\partial \chi_v}{\partial t} = \frac{1}{4\pi\varepsilon_o} \iiint_V \frac{1}{R} \int_{R/v}^{R/c} d\tau \frac{\partial}{\partial t} \rho\left(\vec{r}\,',t-\tau\right) dV' . \quad \text{(A2)}$$

Then integrating with respect to $t$ we find

$$\chi_v\left(\vec{r},t\right) = \frac{-1}{4\pi\varepsilon_o} \iiint_V \frac{1}{R} \int_{R/v}^{R/c} d\tau \rho\left(\vec{r}\,',t-\tau\right) dV' . \quad \text{(A3)}$$

Ultimately, we will need the gradient of (A3) to evaluate the vector potential with the $v$-gauge. To compute the gradient we differentiate only the spatial terms involving $R$, as indicated by the brackets in the equation below,

$$\nabla\chi_v\left(\vec{r},t\right) = -\frac{1}{4\pi\varepsilon_o} \iiint_V \nabla\left(\frac{1}{R} \int_{R/v}^{R/c}\right) d\tau \rho\left(\vec{r}\,',t-\tau\right) dV' . \quad \text{(A4)}$$

Using the product rule and Leibniz's rule along with $\nabla\vec{R} = \hat{R}$ and $\nabla\left(1/R\right) = -\hat{R}/R^2$ we obtain

$$\nabla\chi_v\left(\vec{r},t\right) = \frac{1}{4\pi\varepsilon_o} \iiint_V \left\{ \begin{array}{c} -\dfrac{\hat{R}}{R^2}\,\tau\,\rho\left(\vec{r}\,',t-\tau\right)\Big|_{R/v}^{R/c} \\[2mm] +\dfrac{\hat{R}}{R^2}\displaystyle\int_{R/v}^{R/c} d\tau\,\rho\left(\vec{r}\,',t-\tau\right) \end{array} \right\} dV' . \quad \text{(A5)}$$

Both terms of (A5) can be more compactly written using integration by parts, resulting in

$$\nabla\chi_v\left(\vec{r},t\right) = \frac{-1}{4\pi\varepsilon_o} \iiint_V \frac{\hat{R}}{R^2} \left\{ \int_{R/v}^{R/c} \tau\,\frac{\partial\rho\left(\vec{r}\,',t-\tau\right)}{\partial\tau} d\tau \right\} dV' . \quad \text{(A6)}$$

Using the continuity relation

$$\nabla'\cdot\vec{J}\left(\vec{r},t-\tau\right) = -\frac{\partial\rho\left(\vec{r}\,',t-\tau\right)}{\partial t} = +\frac{\partial\rho\left(\vec{r}\,',t-\tau\right)}{\partial\tau} \quad \text{(A7)}$$

We substitute into (A6) to obtain the SI unit version of Jackson's (7.6) in [1],

$$\nabla\chi_v\left(\vec{r},t\right) = \frac{-1}{4\pi\varepsilon_o} \iiint_V \frac{\hat{R}}{R^2} \int_{R/v}^{R/c} \tau\,d\tau\,\nabla'\cdot\vec{J}\left(\vec{r}\,',t-\tau\right) dV' . \quad \text{(A8)}$$

With the gradient relationship $\nabla' f\left(\vec{R},t\right) = -\nabla f\left(\vec{R},t\right)$ and integration by parts, we obtain the $k^{th}$ Cartesian component of the gradient,

$$\left(\nabla\chi_v\left(\vec{r},t\right)\right)_k = \frac{-1}{4\pi\varepsilon_o} \iiint_V \frac{\partial}{\partial x_k}\left[ \frac{\hat{R}}{R^2} \int_{R/v}^{R/c} d\tau\,\tau\,J_k\left(\vec{r}\,',t-\tau\right) \right] dV' . \quad \text{(A9)}$$

Additional steps lead to the $v$-gauge vector potential given in (12a), which is the SI unit version of (7.7) derived in [1].

## APPENDIX B
### Derivation of Hertz Dipole Potentials

The velocity gauge scalar potential given by the integral in (10) is evaluated as shown below for the case of the Hertz dipole. Integration of the delta functions sifts out the point values of the charges $\pm q(t)$ at the ends of the filament. Contributions of these charges to the potential are delayed by propagation velocity $v$ over the distances $R_1$ and $R_2$ shown in Fig. 2.

$$\phi_v(r,\theta,t) = \frac{1}{4\pi\varepsilon_o} \iiint_V \frac{1}{R} \rho(\vec{r}\,', t - R/v)\, dV'$$

$$= \frac{1}{4\pi\varepsilon_o}\left[\frac{1}{R_1} q(t - R_1/v) - \frac{1}{R_2} q(t - R_2/v)\right]$$

$$= \frac{I_o}{4\pi\varepsilon_o \omega}\left[\frac{1}{R_1}\sin\omega(t - R_1/v) - \frac{1}{R_2}\sin\omega(t - R_2/v)\right] \quad \text{(B1)}$$

$$\doteq \frac{I_o}{4\pi\varepsilon_o \omega}\left[\frac{\sin\omega\left(t - \dfrac{r}{v} + \dfrac{d\ell}{2v}\cos\theta\right)}{r - \dfrac{d\ell}{2}\cos\theta} - \frac{\sin\omega\left(t - \dfrac{r}{v} - \dfrac{d\ell}{2v}\cos\theta\right)}{r + \dfrac{d\ell}{2}\cos\theta}\right]$$

$$= \frac{I_o d\ell \cos\theta}{4\pi\varepsilon_o}\left[\frac{1}{v\,r}\cos\omega(t - r/v) + \frac{1}{\omega r^2}\sin\omega(t - r/v)\right].$$

Note that the second line provides the potential for any balanced time-domain charge density $\pm q(t)$ existing on the ends of the dipole. The fourth line of (B1) results by assuming an infinitesimal dipole length, so that $r >> d\ell$, giving $R_1 \doteq r - (d\ell/2)\cos\theta$ and $R_2 \doteq r + (d\ell/2)\cos\theta$ in spherical coordinates. The final line of (B1) results from using the relationship $\sin(\alpha \pm \beta) = \sin\alpha\cos\beta \pm \cos\alpha\sin\beta$.

The velocity gauge vector potential will be computed by using (26) to transition from the Lorenz gauge in (13). The integral in (13) is easily evaluated using the current density in (22), assuming an infinitesimal dipole length with spatially uniform $I(t)$. Since $r >> d\ell$, the term $1/R(z') \doteq 1/r$ and we can write the Lorenz gauge vector potential as

$$\vec{A}_L(\vec{r},t) = \frac{\mu_o}{4\pi}\iiint_V \frac{1}{R}\vec{J}(\vec{r}\,', t - R/c)\, dV'$$

$$= \frac{\mu_o \hat{z}}{4\pi}\iiint_V \left\{\begin{array}{l}\dfrac{1}{R} I(t - R/c)\delta(x')\delta(y') \bullet \\[2mm] \left[u(z' + d\ell/2) - u(z' - d\ell/2)\right]\end{array}\right\} dx'dy'dz' \quad \text{(B2)}$$

$$= \frac{\mu_o \hat{z}}{4\pi r} I(t - r/c) = \hat{z}\frac{\mu_o I_o d\ell}{4\pi r}\cos\omega(t - r/c).$$

The last line of (B2) first shows the general result for any $I(t)$ followed by that for our example case, where $I(t) = I_o \cos\omega t$.

To transition to the $v$-gauge vector potential using (26) we will also need the gradient of the velocity gauge function.

This is done by first applying (4b), where both $\phi_v(r,\theta,t)$ and $\phi_L(r,\theta,t)$ are obtained from (24), the latter using $v = c$.

$$\frac{\partial \chi_v(r,\theta,t)}{\partial t} = \phi_L(r,\theta,t) - \phi_v(r,\theta,t) =$$

$$\frac{I_o d\ell \cos\theta}{4\pi\varepsilon_o}\left[\begin{array}{l}\dfrac{1}{c\,r}\cos\omega(t - r/c) + \dfrac{1}{\omega r^2}\sin\omega(t - r/c) \\[3mm] -\dfrac{1}{v\,r}\cos\omega(t - r/v) - \dfrac{1}{\omega r^2}\sin\omega(t - r/v)\end{array}\right] \quad \text{(B3)}$$

Next, we integrate (B3) over $t$, with zero integration constant, to obtain the velocity gauge function

$$\chi_v(r,\theta,t) = \frac{I_0 d\ell \cos\theta}{4\pi\varepsilon_o \omega} \bullet$$

$$\left[\begin{array}{l}\dfrac{1}{c\,r}\sin\omega(t - r/c) - \dfrac{1}{\omega r^2}\cos\omega(t - r/c) \\[3mm] -\dfrac{1}{v\,r}\sin\omega(t - r/v) + \dfrac{1}{\omega r^2}\cos\omega(t - r/v)\end{array}\right]. \quad \text{(B4)}$$

Finally, the gradient operation is performed in spherical coordinates, noting no $\varphi$ dependence, to yield

$$\nabla\chi_v(r,\theta,t) = \hat{r}\frac{\partial\chi_v}{\partial r} + \hat{\theta}\frac{1}{r}\frac{\partial\chi_v}{\partial\theta} = \frac{I_o d\ell}{4\pi\varepsilon_o} \bullet$$

$$\left\{\begin{array}{l}\hat{r}\cos\theta\left[\begin{array}{l}-\dfrac{2}{c^2 r}\sin\omega(t - r/c) - \dfrac{\omega}{c^2 r}\cos\omega(t - r/c) \\[2mm] +\dfrac{2}{\omega r^3}\cos\omega(t - r/c) + \dfrac{2}{v\,r^2}\sin\omega(t - r/v) \\[2mm] +\dfrac{\omega}{v^2 r}\cos\omega(t - r/v) - \dfrac{2}{\omega r^3}\cos\omega(t - r/v)\end{array}\right] \\[12mm] -\hat{\theta}\sin\theta\left[\begin{array}{l}\dfrac{1}{c\,r^2}\sin\omega(t - r/c) - \dfrac{1}{\omega r^3}\cos\omega(t - r/c) \\[2mm] -\dfrac{1}{v\,r^2}\sin\omega(t - r/v) + \dfrac{1}{\omega r^3}\cos\omega(t - r/v)\end{array}\right]\end{array}\right\}. \quad \text{(B5)}$$

Adding (B5) to the spherical components of $\vec{A}_L(\vec{r},t)$ given in (B2), namely $A_r = A_z\cos\theta$ and $A_\theta = -A_z\sin\theta$, yields the final expression for the $v$-gauge vector potential shown in equations (29a), (29b) and (29c).

**ACKNOWLEDGMENTS**

The authors are thankful to Dr. Kelvin Lee and to Prof. Robert G. Olsen, (Emeritus), Washington State University for valuable discussions during the preparation of this paper. We are also grateful to Prof. Robert D. Nevels, Texas A&M University, for reading our draft manuscript and suggesting many improvements which have been made.